\documentstyle[12pt]{article}
\begin{document}
\noindent
\begin{center}
{\Large {\bf Local Gravity Constraints and\\ Power Law f(R)
Theories\\}} \vspace{2cm}

 ${\bf Yousef~
Bisabr}$\footnote{e-mail:~y-bisabr@srttu.edu.}\\
\vspace{.5cm} {\small{Department of Physics, Shahid Rajaee Teacher
Training University,
Lavizan, Tehran 16788, Iran.}}\\
\end{center}
\vspace{1cm}
\begin{abstract}
There is a conformal equivalence between power law $f(R)$ theories
and scalar field theories in which the scalar degree of freedom
evolves under the action of an exponential potential function. In
the scalar field representation there is a strong coupling of the
scalar field with the matter sector due to the conformal
transformation. We use chameleon mechanism to implement
constraints on the potential function of the scalar field in order
that the resulting model be in accord with Solar System
experiments.  Investigation of these constraints reveals that
there may be no possibility to distinguish between a power law
$f(R)$ function and the usual Einstein-Hilbert Lagrangian density.

\end{abstract}
 \vspace{3cm}

There are strong observational evidences that the expansion of the
universe is accelerating. These observations are based on type Ia
supernova \cite{super}, cosmic microwave background radiation
\cite{cmbr}, large scale structure formation \cite{ls}, weak
lensing \cite{wl}, etc. The standard explanation invokes an
unknown component, usually referred to as dark energy. It
contributes to energy density of the universe with
$\Omega_{d}=0.7$ where $\Omega_{d}$ is the corresponding density
parameter, see e.g., \cite{ca} and references therein. The
simplest dark energy scenario which seems to be both natural and
consistent with observations is the $\Lambda$CDM model in which
dark energy is identified as a cosmological constant \cite{ca}
\cite{wei} \cite{cc}. However, in order to avoid theoretical
problems \cite{wei}, other scenarios have been investigated. Among
these scenarios, there are quintessence \cite{q}, tachyons
\cite{tach}, phantom \cite{ph}, quintom \cite{qui} and modified
gravity models \cite{modi}.  In the latter, one modifies the laws
of gravity whereby a late time acceleration  is produced without
recourse to a dark energy component.  One family of these modified
gravity models is obtained by replacing the Ricci scalar $R$ in
the usual Einstein-Hilbert Lagrangian density for some function
$f(R)$. However, changing gravity Lagrangian have consequences not
only in cosmological scales but also in galactic ones so that it
seems to be necessary to investigate the low energy limit of such
$f(R)$ theories.\\
Early works on the weak field limit of $f(R)$ theories led to
negative results.  In fact, using the equivalence of $f(R)$ and
scalar-tensor theories \cite{soko} \cite{maeda} \cite{wands}, it
is originally suggested that all $f(R)$ theories should be ruled
out \cite{chiba} since they violate the weak field constraints
coming from Solar System experiments. This claim was based on the
fact that $f(R)$ theories (in the metric formalism) are equivalent
to Brans-Dicke theory with $\omega=0$ while observations set the
constraint $\omega>40000$ \cite{will}.  In this case the
post-Newtonian parameter satisfies $\gamma=\frac{1}{2}$ instead of
being equal to unity as required by observations.  Later, it was
noted by many authors that for scalar fields with sufficiently
large mass it is possible to drive $\gamma$ close to unity even
for null Brans-Dicke parameter.  In this case the scalar field
becomes short-ranged and has no effect at Solar System scales.
Recently, it is shown that there exists an important possibility
that the effective mass of the scalar field be scale dependent
\cite{k}. In this chameleon mechanism, the scalar field may
acquire a large effective mass in Solar System scale so that it
hides local experiments while at cosmological scales it is
effectively light and may provide an appropriate cosmological behavior.\\
In the present work we intend to use this criterion to set local
gravity constraints on $f(R)$ theories.  There are a number of
works concerning these constraints on $f(R)$ theories \cite{f1}
\cite{f2}. We will focus on power law $f(R)$ theories and show
that the constraints on the parameters space suggest that they are
hardly
distinguishable from $\Lambda$CDM scenario.\\\\
Let us begin with the following action\footnote{We use the unit
$(8\pi G)^{-1}=1.$}
\begin{equation}
S=\frac{1}{2} \int d^{4}x \sqrt{-g}~ f(R) + S_{m}(g_{\mu\nu},
\psi)\label{a}\end{equation} where $g$ is the determinant of
$g_{\mu\nu}$, $f(R)$ is an unknown function of the scalar
curvature $R$ and $S_{m}$ is the matter action depending on the
metric $g_{\mu\nu}$ and some matter field $\psi$. We may use a new
set of variables
\begin{equation}
\bar{g}_{\mu\nu} =p~ g_{\mu\nu} \label{c}\end{equation}
\label{a1}\begin{equation} \phi = \frac{1}{2\beta} \ln p
\label{dd}\end{equation}
 where
$p\equiv\frac{df}{dR}=f^{'}(R)$ and $\beta=\sqrt{\frac{1}{6}}$.
This is indeed a conformal transformation which transforms the
above action in the Jordan frame to the Einstein frame \cite{soko}
\cite{maeda} \cite{wands}
\begin{equation}
S=\frac{1}{2} \int d^{4}x \sqrt{-g}~\{ \bar{R}-\bar{g}^{\mu\nu}
\partial_{\mu} \phi~ \partial_{\nu} \phi -2V(\phi)\} + S_{m}(\bar{g}_{\mu\nu}
e^{2\beta \phi}, \psi) \label{s}\end{equation} In the Einstein
frame, $\phi$ is a scalar field with a self-interacting potential
which is given by
\begin{equation}
V(\phi)=\frac{1}{2}e^{-2\beta \phi} \{r[p(\phi)]-e^{-2\beta \phi}
f(r[p(\phi)])\} \label{v}\end{equation} where $r(p)$ is a solution
of the equation $f^{'}[r(p)]-p=0$ \cite{soko}.  One usually states
that any $f(R)$ gravity model is mathematically equivalent with a
minimally coupled scalar field theory with an appropriate
potential function.  In general, this does not imply physical
equivalence of the two conformal frames. In fact it is shown that
some physical systems can be differently interpreted in different
conformal frames \cite {soko} \cite{no}.  The physical status of
the two conformal frames is an open question which we are not
going to address here.  However we assume that the scalar degree
of freedom in the Einstein frame should satisfy stringent
constraints from solar system experiments.  It is important to
note that conformal transformation induces the coupling of the
scalar field $\phi$ with the matter sector. The strength of this
coupling $\beta$, is fixed to be $\sqrt{\frac{1}{6}}$ and is the
same for all types of matter fields.  In the case of such a strong
matter coupling, the role of the potential of the scalar field is
important for consistency with local gravity experiments. When the
potential satisfies certain conditions it is possible to attribute
an effective mass to the scalar field which has a strong
dependence on ambient density of matter.  A theory in which such a
dependence is realized is said to be a chameleon theory \cite{k}.
In such a theory the scalar field $\phi$ can be heavy enough in
the environment of the laboratory tests so that the local gravity
constraints suppressed even if $\beta$ is of the order of unity.
Meanwhile, it can be light enough in the low-density cosmological
environment to be
considered as a candidate for dark energy.\\
Variation of the action (\ref{a}) with respect to
$\bar{g}_{\mu\nu}$ and $\phi$, gives the field equations
\begin{equation}
\bar{G}_{\mu\nu}=\partial_{\mu} \phi~\partial_{\nu} \phi
-\frac{1}{2}\bar{g}_{\mu\nu} \partial_{\gamma}
\phi~\partial^{\gamma} \phi-V(\phi)
\bar{g}_{\mu\nu}+\bar{T}_{\mu\nu} \label{g}\end{equation}
\begin{equation}
\bar{\Box} \phi -\frac{dV}{d\phi}=-\beta \bar{T}
\label{b}\end{equation} where
\begin{equation}
\bar{T}_{\mu\nu}=\frac{-2}{\sqrt{-g}}\frac{\delta S_{m}}{\delta
\bar{g}^{\mu\nu}}\label{t}\end{equation} and
$\bar{T}=\bar{g}^{\mu\nu}\bar{T}_{\mu\nu}$.  Covariant
differentiation of (\ref{g}) and the Bianchi identities give
\begin{equation}
\bar{\nabla}^{\mu} \bar{T}_{\mu\nu}=\beta~\bar{T}~\partial_{\nu}
\phi \label{b1}\end{equation} which implies that the matter field
is not generally conserved and feels a new force due to gradient
of the scalar field.  Let us consider $\bar{T}_{\mu\nu}$ as the
stress-tensor of dust with energy density $\bar{\rho}$ in the
Einstein frame. In a static and spherically symmetric spacetime
the equation (\ref{b}) gives
\begin{equation}
\frac{d^2 \phi}{d\bar{r}^2}+\frac{2}{\bar{r}}
\frac{d\phi}{d\bar{r}}=\frac{dV_{eff}(\phi)}{d\phi}
\label{d}\end{equation} where $\bar{r}$ is distance from center of
the symmetry in the Einstein frame and
\begin{equation}
V_{eff}(\phi)=V(\phi)-\frac{1}{4}\rho e^{-4\beta \phi}
\label{d1}\end{equation} Here we have used the relation
$\bar{\rho}=e^{-4\beta \phi} \rho$ that relates the energy
densities in the Jordan and the Einstein frames.  We consider a
spherically symmetric body with a radius $\bar{r}_{c}$ and a
constant energy density $\bar{\rho}_{in}$ ($\bar{r}<
\bar{r}_{c}$). We also assume that the energy density outside the
body ($\bar{r}> \bar{r}_{c}$) is given by $\bar{\rho}_{out}$. We
will denote by $\varphi_{in}$ and $\varphi_{out}$ the field values
at two minima of the effective potential $V_{eff}(\phi)$ inside
and outside the object, respectively. They must clearly satisfy
$V^{'}_{eff}(\varphi_{in})=0$ and $V^{'}_{eff}(\varphi_{out})=0$
where prime indicates differentiation of $V_{eff}(\phi)$ with
respect to the argument.  As usual, masses of small fluctuations
about these minima are given by
$m_{in}=[V^{''}_{eff}(\varphi_{in})]^{\frac{1}{2}}$ and
$m_{out}=[V^{''}_{eff}(\varphi_{out})]^{\frac{1}{2}}$ which depend
on ambient matter density. A region with large mass density
corresponds to a heavy mass field while regions with low mass
density corresponds to a field with lighter mass. In this way it
is possible for the mass field to take sufficiently large values
near massive objects in the Solar System scale and to hide the
local tests. For a spherically symmetric body there is a
thin-shell condition
\begin{equation}
\frac{\Delta
\bar{r}_{c}}{\bar{r}_{c}}=\frac{\varphi_{out}-\varphi_{in}}{6\beta
\Phi_{c}}\ll 1 \label{l1}\end{equation} where $\Phi_{c}$ is the
Newtonian potential at $\bar{r}=\bar{r}_{c}$. In this case,
equation (\ref{d}) with some appropriate boundary conditions gives
the field profile outside the object \cite{k}
\begin{equation}
\phi(\bar{r})=-\frac{\beta}{4\pi} \frac{3\Delta
\bar{r}_{c}}{\bar{r}_{c}}\frac{M_{c} e^{-m_{out}
(\bar{r}-\bar{r}_{c})}}{\bar{r}}+\varphi_{out}
\label{l2}\end{equation} where $M_{c}$ is mass of the object.
\\The function $f(R)$ in the Jordan frame is closely related to the
potential function of the scalar degree of freedom of the theory
in the Einstein frame.  Any functional form for the potential
function corresponds to a particular class of $f(R)$ theories. To
find a viable function $f(R)$ passing Solar System tests one can
equivalently work with its corresponding potential function in the
Einstein frame and put constraints on the relevant parameters via
chameleon mechanism. Taking this as our criterion, we will use a
pure exponential potential function. There are two reasons for
this choice.  Firstly, this class of potentials arises in a number
of physical situations.  In particular, there are reports that
quintessence field with exponential potentials can produce late
time acceleration \cite{w}.  Secondly, as we will show in the
following an exponential potential function for the scalar field
corresponds to a power law $f(R)$ theory.  In fact, there are
arguments concerning cosmological viability of this class of
$f(R)$ models \cite{car} \cite{am}.  Moreover, it is argued that
some power law $f(R)$ theories may have sting/M-theory origin
\cite{sm}.  These arguments make investigation of viability of
these models in terms of local experiments be a mandate.  \\To
have a pure exponential potential function, we take
\begin{equation}
r(p)=[p~\frac{n}{\alpha~(n+1)}]^{n}
 \label{r}\end{equation} where $\alpha$ and
$n$ are constant parameters.  It then leads to
\begin{equation}
V(\phi)=\frac{1}{2(n+1)}[\alpha~\frac{n+1}{n}]^{-n}
~e^{2(n-1)\beta \phi} \label{r1}\end{equation}   On the other
hand, this choice of the $r(p)$  function gives a power law $f(R)$
theory
\begin{equation}
f(R)=\alpha~R^{1+\frac{1}{n}} \label{r2}\end{equation} Thus there
is a correspondence between a power law $f(R)$ theory in the
Jordan frame and a minimally coupled scalar field theory with an
exponential potential in the Einstein frame\footnote{It should be
remarked that in the end of the present work we became aware of
\cite{capoz} in which a relation between power law f(R) theories
and a minimally coupled scalar field with specific exponential
potential has been reported in the absence of matter systems.}. We
can now find the solution of $V^{'}_{eff}(\phi)=0$ by substituting
(\ref{r1}) into (\ref{d1})
\begin{equation}
\varphi=\frac{1}{2\beta (n+1)}\ln \{\rho~\frac{n+1}{1-n}~(\alpha
\frac{n+1}{n})^{n}\}\label{n} \label{v}\end{equation} In order
that $\varphi$ be a local minimum we should have
\begin{equation}
\frac{n+1}{1-n}~(\alpha \frac{n+1}{n})^{n}>0
\label{c}\end{equation}   In the following we shall consider
thin-shell condition and
the constraints set by equivalence principle and fifth force experiments.\\\\
$1.~ Thin-shell~~ condition$\\
In the chameleon mechanism, the chameleon field is trapped inside
large and massive bodies and its influence on the other bodies is
only due to a thin-shell near the surface of the body.  The
criterion for this thin-shell condition is given by (\ref{l1}). If
we combine (\ref{l1}) and (\ref{n}) we obtain
\begin{equation}
\frac{\Delta \bar{r}_{c}}{\bar{r}_{c}}=\frac{1}{12\beta^2 (n+1)}
\frac{1}{\Phi_{c}} \ln \frac{\rho_{out}}{\rho_{in}}
\label{h}\end{equation} where $\rho_{in}$ and $\rho_{out}$ are
energy densities inside and outside of the body in the Jordan
frame, respectively, and $\Phi_{c}=M_{c}/8\pi \bar{r}_{c}$ with
$M_{c}$ being the mass of the body. In weak field approximation
the spherically symmetric metric in the Jordan frame is given by
\begin{equation}
ds^{2}=-[1-2X(r)]dt^{2}+[1+2Y(r)]dr^{2}+r^2 d\Omega^2
\end{equation}
where $X(r)$ and $Y(r)$ are some functions of $r$. There is a
relation between $r$ and $\bar{r}$ so that $\bar{r}=p^{1/2} r$. If
we consider this relation under the assumption $ m_{out}~r\ll 1 $,
namely that the Compton wavelength $m_{out}^{-1}$ is much larger
than Solar System scales, then we have $\bar{r} \approx r$. In
this case, the chameleon mechanism gives for the post-Newtonian
parameter $\gamma$ \cite{faulk}
\begin{equation}
\gamma=\frac{3-\frac{\Delta r_{c}}{r_{c}}}{3+\frac{\Delta
r_{c}}{r_{c}}} \simeq 1-\frac{2}{3}\frac{\Delta r_{c}}{r_{c}}
\label{gg}\end{equation}\\
We can now apply (\ref{h}) on the Earth and obtain the condition
that the Earth has a thin-shell. To do this, we assume that the
Earth is a solid sphere of radius $R_{e} =6.4\times 10^{8}~cm$ and
mean density $\rho_{e} \sim 10~gr/cm^{3}$. We also assume that the
Earth is surrounded by an atmosphere with homogenous density
$\rho_{a} \sim 10^{-3}~gr/cm^{3}$ and thickness $100 km$.  Then we
rewrite equation (\ref{h})
\begin{equation}
\frac{\Delta R_{e}}{R_{e}}=\frac{1}{12\beta^2 (n+1)}
\frac{1}{\Phi_{e}} \ln \frac{\rho_{a}}{\rho_{e}}
\label{hhh}\end{equation} With $\Phi_{e}=6.95\times 10^{-10}$
\cite{wein}, Newtonian potential on surface of the Earth, it gives
\begin{equation}
\frac{\Delta R_{e}}{R_{e}}=-\frac{4.96\times 10^{9}}{n+1}
\label{rrr}\end{equation} The tightest Solar System constraint on
$\gamma$ comes from Cassini tracking which gives $\mid \gamma -1
\mid < 2.3 \times 10^{-5}$ \cite{will}.  This together with
(\ref{gg}) and (\ref{rrr}) yields an upper bound for the parameter
$n$
\begin{equation} |n+1|>1.44 \times 10^{14}
\label{n+1}\end{equation} Combining this result with (\ref{r2})
reveals that this power law $f(R)$ theory hardly deviates from
general relativity.\\\\
2. $Equivalence ~ principle$\\
We now consider constraints coming from possible violation of weak
equivalence principle. We assume that the Earth, together with its
surrounding atmosphere, is an isolated body and neglect the effect
of the other compact objects such as the Sun, the Moon and the
other planets. Far away the Earth, matter density is modeled by a
homogeneous gas with energy density $\rho_{G}\sim 10^{-24}
gr/cm^{3}$.  To proceed further, we first consider the condition
that the atmosphere of the Earth satisfies the thin-shell
condition \cite{k}. If the atmosphere has a thin-shell the
thickness of the shell ($\Delta R_{a}$) must be clearly smaller
than that of the atmosphere itself, namely $\Delta R_{a}< R_{a}$,
where $R_{a}$ is the outer radius of the atmosphere. If we take
thickness of the shell equal to that of the atmosphere itself
$\Delta R_{a}\sim 10^{2}~km$ we obtain $\frac{\Delta
R_{a}}{R_{a}}<1.5\times 10^{-2}$. It is then possible to relate
$\frac{\Delta R_{e}}{R_{e}}=\frac{\varphi_{a}-\varphi_{e}}{6\beta
\Phi_{e}}$ and $\frac{\Delta
R_{a}}{R_{a}}=\frac{\varphi_{G}-\varphi_{a}}{6\beta \Phi_{a}}$
where $\varphi_{e}$, $\varphi_{a}$ and $\varphi_{G}$ are the field
values at the local minimum of the effective potential in the
regions $r<R_{e}$ , $R_{a}>r>R_{e}$ and $r>R_{a}$ respectively.
Using the fact that newtonian potential inside a spherically
symmetric object with mass density $\rho$ is $\Phi \propto\rho
R^{2}$, one can write $\Phi_{e}=10^{4}~\Phi_{a}$ where $\Phi_{e}$
and $\Phi_{a}$ are Newtonian potentials on the surface of the
Earth and the atmosphere, respectively. This gives $\Delta R_{e}/
R_{e} \approx 10^{-4}~\Delta R_{a}/ R_{a}$. With these results,
the condition for the atmosphere to have a thin-shell is
\begin{equation}
\frac{\Delta R_{e}}{R_{e}} < 1.5\times 10^{-6}
\label{R}\end{equation}\\
The tests of equivalence principle measure the difference of
free-fall acceleration of the Moon and the Earth towards the Sun.
The constraint on the difference of the two acceleration is given
by \cite{will}
\begin{equation}
\frac{|a_{m}-a_{e}|}{a_{N}} < 10^{-13} \label{f}\end{equation}
where $a_{m}$ and $a_{e}$ are acceleration of the Moon and the
Earth respectively and $a_{N}$ is the Newtonian acceleration.  The
Sun and the Moon are all subject to the thin-shell condition
\cite{k} and the field profile outside the spheres are given by
(\ref{l2}) with replacement of corresponding quantities.  The
accelerations $a_{m}$ and $a_{e}$ are then given by \cite{k}
\begin{equation}
a_{e}\approx a_{N}\{1+18\beta^2 (\frac{\Delta R_{e}}{R_{e}})^2
\frac{\Phi_{e}}{\Phi_{s}}\}
\end{equation}
\begin{equation}
a_{m}\approx a_{N}\{1+18\beta^2 (\frac{\Delta R_{e}}{R_{e}})^2
\frac{\Phi_{e}^2}{\Phi_{s}\Phi_{m}}\}
\end{equation}
where $\Phi_{e}=6.95\times 10^{-10}$, $\Phi_{m}=3.14\times
10^{-11}$ and $\Phi_{s}=2.12\times 10^{-6}$ are Newtonian
potentials on the surfaces of the Earth, the Moon and the Sun,
respectively \cite{wein}. This gives a difference of free-fall
acceleration
\begin{equation}
\frac{|a_{m}-a_{e}|}{a_{N}}=(0.13)~ \beta^2~(\frac{\Delta
R_{e}}{R_{e}})^2
\end{equation}
Combining this with (\ref{f}) results in
\begin{equation}
\frac{\Delta R_{e}}{R_{e}} < 6.74\times 10^{-6}
\label{RR}\end{equation} which is of the same order of the
condition (\ref{R}) that the atmosphere has a thin-shell.  Taking
this as the constraint coming from violation of equivalence
principle, we obtain
\begin{equation}
\mid n+1 \mid> 1.67\times 10^{15} \label{nn}\end{equation} which
is not much different from the bound given by
(\ref{n+1}).\\\\
3. $Fifth~force$\\
The potential energy associated with a fifth force interaction is
parameterized by a Yukawa-type potential
\begin{equation}
U(r)=-\alpha\frac{m_{1}m_{2}}{8\pi}\frac{e^{-r/\lambda}}{r}
\label{y}\end{equation} where $m_{1}$ and $m_{2}$ are masses of
the two test bodies separating by distance r, $\alpha$ is strength
of the interaction and $\lambda$ is the range.  Thus fifth force
experiment constrains regions of ($\alpha, \lambda$) parameter
space. These experiments are usually carried out in a vacuum
chamber in which the range of the interaction inside it is of the
order of the size of the chamber \cite{k}, namely $\lambda\sim
R_{vac}$. The tightest bound on the strength of the interaction is
$\alpha<10^{-3}$ \cite{h}.  Inside the chamber we consider two
identical bodies with uniform densities $\rho_{c}$, radii $r_{c}$
and masses $m_{c}$.  If the two bodies satisfy the thin-shell
condition, their field profile outside the bodies are given by
\begin{equation}
\phi(r)=-\frac{\beta}{4\pi} \frac{3\Delta r_{c}}{r_{c}}\frac{m_{c}
~e^{-r/R_{vac}}}{r}+\varphi_{vac} \label{22}\end{equation} Then
the corresponding potential energy of the interaction is
\begin{equation}
V(r)=-2\beta^2 (\frac{3\Delta
r_{c}}{r_{c}})^{2}~\frac{m_{c}^2}{8\pi}\frac{e^{-r/R_{vac}}}{r}
\label{yy}\end{equation} The bound on the strength of the
interaction translates into
\begin{equation}
2\beta^2 (\frac{3\Delta r_{c}}{r_{c}})^{2}<10^{-3}
\label{222}\end{equation} One can write for each of the test
bodies
\begin{equation}
\frac{\Delta r_{c}}{r_{c}}=\frac{1}{12\beta^2 (n+1)}
\frac{1}{\Phi_{c}} \ln \frac{\rho_{vac}}{\rho_{c}}
\label{hh}\end{equation} where $\rho_{vac}$ is energy density of
the vacuum inside the chamber.  In the experiment carried out in
\cite{h}, one used a typical test body with mass $m_{c}\approx 40
gr$ and radius $r_{c}\approx 1 cm$. These correspond to $\rho_{c}
\approx 9.5 gr/cm^{3}$ and $\Phi_{c} \sim 10^{-27}$. Moreover, the
pressure in the vacuum chamber was reported to be $3\times
10^{-8}$ Torr which is equivalent to $\rho_{vac} \approx 4.8\times
10^{-14}~gr/cm^{3}$\footnote{Note that due to the logarithmic
dependence of $\frac{\Delta r_{c}}{r_{c}}$ with $\rho_{vac}$ the
value of the vacuum energy density does not effectively change the
order of magnitude of (\ref{21}).}. Substituting these into
(\ref{hh}) and combining the result with (\ref{222}) gives the
bound
\begin{equation}
\mid n+1 \mid > 3\times 10^{29} \label{21}\end{equation} which is
much stronger than (\ref{n+1}) and (\ref{nn}).  As the last point,
there are some remarks to do with respect to stability of the
model (\ref{r2}).  In principle, stability issues should be
considered to make sure that an $f(R)$ model is viable \cite{st}.
In particular, stability in matter sector (the Dolgov-Kawasaki
instability \cite{dk}) imposes some conditions on the functional
form of $f(R)$ models.  The first theories which easily pass this
instability have been presented in \cite{f1} \cite{stab}.  These
conditions require that the first and the second derivatives of
$f(R)$ function with respect to the Ricci scalar $R$ should be
positive definite.  The positivity of the first derivative ensures
that the scalar degree of freedom is not tachyonic and positivity
of the second derivative tells us that graviton is not a ghost.
For power law $f(R)$ theories of the type (\ref{r2}), we should
have $n>-1$ and $n>0$ to ensure that
$f'(R)>0$ and $f''(R)>0$, respectively.\\
In summary, we have discussed viability of power law $f(R)$
theories in terms of local gravity constraints.  We have used the
correspondence between this class of $f(R)$ theories to scalar
field theories with an exponential self-interacting potential.  In
the scalar field representation of the theory there is a strong
coupling of the scalar field with the matter sector. We have
considered the conditions that this coupling is suppressed by
chameleon mechanism.  We have found that in order that the theory
be consistent with local gravity experiments the exponent of the
curvature scalar hardly deviates from unity. The constraint
(\ref{21}) is much stronger than that reported in \cite{clif}
which is obtained by considering the perihelion precession of
Mercury under the assumption that it follows timelike geodesics.
Our results preclude the possibility of regarding power law $f(R)$
models as
 viable candidates for generalizing general relativity.

\end{document}